\documentclass[12pt]{article}
\usepackage[margin=1in]{geometry}
\usepackage{amsthm,amssymb,mathrsfs,setspace,amsmath,latexsym,footmisc,graphicx,booktabs}
\usepackage{color}
\usepackage{textcomp}
\usepackage{multirow} 
\usepackage{makecell}
\usepackage{subfig}
\usepackage{enumitem}

\theoremstyle{plain}

\theoremstyle{definition}

\newcommand{\btheta}{\boldsymbol{\theta}}

\theoremstyle{remark}

\begin{document}
	
\title{Application of a General Family of Bivariate Distributions in Modelling Dependent Competing Risks Data with Associated Model Selection}

\author{Aakash Agrawal \thanks{Independent Researcher, Email: akash2016@alumni.iitg.ac.in} ,
Ayon Ganguly\thanks{Indian Institute of Technology Guwahati, Assam, India;
Email: aganguly@iitg.ac.in} , and 
Debanjan Mitra \thanks{Indian Institute of Management Udaipur, Rajasthan,
India; Email: debanjan.mitra@iimu.ac.in} }

\date{}	

\maketitle

\begin{abstract}
In this article, a general family of bivariate distributions is used to
model competing risks data with dependent factors. The general structure of
competing risks data considered here includes ties. A comprehensive inferential
framework for the proposed model is presented: maximum likelihood
estimation, confidence interval construction, and model selection within
the bivariate family of distributions for a given dependent competing risks
data. The inferential methods are very convenient to implement. Through
detailed simulations, the inferential methods are observed to provide quite
reasonable results. Analysis of a real data from the Diabetic Retinopathy
Study is carried out with the help of the proposed model as an illustrative
example.  
\end{abstract}


\section{Introduction}
In scenarios with competing risks, there may be multiple causes for the occurrence of the event of interest for subjects under study, and occurrence of any one of the risk factors precludes the occurrence of the other relevant risk factors~\cite{Klein}. Models for competing risks are widely studied in medicine, engineering, finance, etc. Crowder\cite{Crowder} gives a detailed account of classical models on this topic. 

More often than not, in competing risks scenarios, the risk factors influence each other. In medical studies, for example, dependent risks factors are commonly observed in cancer studies where the death of a subject may be the event of interest. Generally speaking, for a subject, death may occur due to the advancement of the particular type of cancer being studied or some other cause(s) that are directly related to cancer, such as the treatment of cancer or the complications that may arise thereafter. For instance, patients in an advanced stage of cancer may be treated with chemotherapy. Some particular drugs used in chemotherapy may have serious adverse effects - at least on a small to moderate percentage of patients treated - such as reduction in the number of white blood cells, which is a life threatening condition by itself.


The earlier statistical literature on competing risks mostly assumed independence among the risk factors (\cite{Cox}, \cite{Crowder}). Under the assumption of independence or latent failure time model~\cite{Cox}, different univariate probability distributions have been used to model lifetimes corresponding to the risk factors. A naturally appealing approach for analyzing dependent competing risks data is, therefore, to consider a suitable bivariate or multivariate probability distribution for jointly modelling the lifetimes corresponding to the risk factors. Copula-based models are commonly used for such purposes, in which lifetime distributions corresponding to the dependent risk factors are connected by an assumed copula from one of the copula families (\cite{Escarela}, \cite{Lo}). 

Lawless~\cite{Lawless} developed the likelihood function for directly using
a bivariate probability distribution as a model for competing risks data.
Several researchers have used this approach very recently. Feizjavadian and
Hashemi~\cite{Feiz2015} have used the Marshall-Olkin bivariate Weibull
(MOBW) distribution~\cite{Marshall67} to model dependent competing risks
data under a progressive hybrid censoring scheme. Samanta and
Kundu~\cite{Samanta} have discussed Bayesian inference for the same model
by using flexible Gamma-Dirichlet prior assumptions. Alqallaf and
Kundu~\cite{Alqallaf} have used a bivariate inverse generalized exponential
distribution for the same problem also. In a different direction, Bai et
al.~\cite{Bai} have used different forms of bivariate exponential
distributions, including the one proposed by Marshall and
Olkin~\cite{Marshall67} to model dependent competing risks data in the
context of step-stress experiments. 

In parametric modelling, it is of utmost importance to select an
appropriate model for a given data, as otherwise, the subsequent inference
is invalid~\cite{Leeb}. In principle, classical goodness-of-fit tests can
check the validity of a fitted model. However, distribution-specific tests, which have more power compared to the omnibus-type tests such as Kolmogorov-Smirnov or Cramer-von Mises, can be quite difficult to develop exploiting properties of the concerned distributions, especially for complicated probability distributions. A relatively simpler solution to the problem of selecting an appropriate model is to use a general parsimonious family of distributions for modelling purposes and then to choose the model which is most appropriate for the given data within the family. This approach has been used by many researchers. For example, generalized gamma distribution, which is a family containing well-known models such as gamma, Weibull, lognormal, and positive stable distributions, has been used as the frailty distribution, followed by a model selection approach~\cite{Bala}. In the context of modelling populations with a cure fraction, i.e., the so-called cure rate models, the Conway-Maxwell Poisson, and the generalized gamma distributions have been used to model the random number of competing causes and lifetimes, respectively; see Balakrishnan and Pal~\cite{Bala-Suvra} and the references therein. 

In this article, we propose to model dependent competing risks data by using a general family of bivariate distributions. The bivariate family is constructed by using the well-known univariate family, called the Lehmann family or the frailty parameter family~\cite{Marshall-book}, following the approach of Marshall and Olkin~\cite{Marshall67} who constructed the MOBW distribution assuming a shock model. The univariate Lehmann family contains Weibull, Gompertz, and Lomax distributions as members of the family, all of which are quite well-known in the context of lifetime data. Therefore, constructing a bivariate family of distributions using the Lehmann family, we arrive at a bivariate family that has a bivariate Weibull, a bivariate Gompertz, and a bivariate Lomax distribution as its members. In fact, the MOBW distribution is a member of this bivariate family. 

The primary advantage of using this general family of bivariate distributions to model dependent competing risks data is its flexibility derived from its member distributions. We develop likelihood inference for this problem and observe that computation of maximum likelihood estimates for the parameters is very convenient in this case. For a given competing risks data with dependent factors, it is naturally of interest to determine the most appropriate model. We carry out a study of model selection within the bivariate family and observe that a simple likelihood-based approach is effective in choosing the suitable model for a given data. The inferential framework presented in this paper is thus quite comprehensive. Moreover, this work generalizes the works of researchers who have used a specific bivariate distribution to model dependent competing risks data. These are the main contributions of this paper. 

The paper is organized as follows. The details of the construction of the bivariate family of distributions are given in Section \ref{sec:biv-model}. Likelihood inference for modelling dependent competing risks data by using the bivariate family is discussed in Section \ref{sec:model-inference}. This section also presents the construction of confidence intervals for model parameters by using approaches such as Fisher information matrix and parametric bootstrap. A likelihood-based approach for model selection is presented in this section as well. In Section \ref{sec:sim}, the results and discussions of a detailed Monte Carlo simulation study are presented. The Monte Carlo study examines the performance of the maximum likelihood estimates (MLEs), confidence intervals, and the model selection approach. Analysis of a real dataset is presented in Section \ref{sec:real-data}. Finally, concluding remarks are made in Section \ref{sec:concl}.  
              
\section{A General Family of Bivariate Distributions} \label{sec:biv-model}
\subsection{The Lehmann family of distributions}
The survival function of the Lehmann family, which is also known as the frailty parameter family~\cite{Marshall-book}, is given by
\begin{equation}
S(t;\alpha, \lambda) = \left( S_0(t; \lambda) \right)^{\alpha},\quad t>0, \label{lehmann}
\end{equation}
where $S_0(\cdot; \lambda)$ is the baseline survival function depending
only on the parameter $\lambda\,(> 0)$. Here, the power parameter
$\alpha\,(> 0)$ is sometimes called the frailty
parameter~\cite{Marshall-book}. Here, we assume that the baseline survival
function $S_0(\cdot;\,\lambda)$ is absolutely continuous.  The hazard rate
function of the Lehmann family is given by
\begin{equation}
h(t;\alpha, \lambda) = \alpha h_0(t;\lambda), \nonumber
\end{equation}
where $h_0(t;\lambda) = -\frac{d}{dt}\log S_0(t:\lambda)$ is the hazard
rate corresponding to the baseline distribution.  

Special members of Lehmann family are Weibull, Gompertz, and Lomax distributions which are obtained for different choices of the baseline survival function $S_0(t; \lambda)$ in Eq.\eqref{lehmann}. In particular, Weibull distribution is obtained when
\begin{equation}
S_0(t; \lambda) = e^{-t^{\lambda}}, \quad t>0. \nonumber
\end{equation}
Gompertz distribution is obtained by choosing
\begin{equation}
S_0(t; \lambda) = e^{-(\exp(\lambda t) - 1)}, \quad t>0, \nonumber
\end{equation}
and for Lomax distribution one chooses 
\begin{equation}
S_0(t; \lambda) = \frac{\lambda t}{1 + \lambda t}, \quad t>0. \nonumber
\end{equation}
With Weibull, Gompertz, and Lomax models - all of which are well-known life distributions - as its members, the Lehmann family is naturally of interest in lifetime data analysis.   
\subsection{Construction of the bivariate family of distributions}
\begin{figure}
\centering
\subfloat{\includegraphics[width=.4\linewidth, height=.4\linewidth]{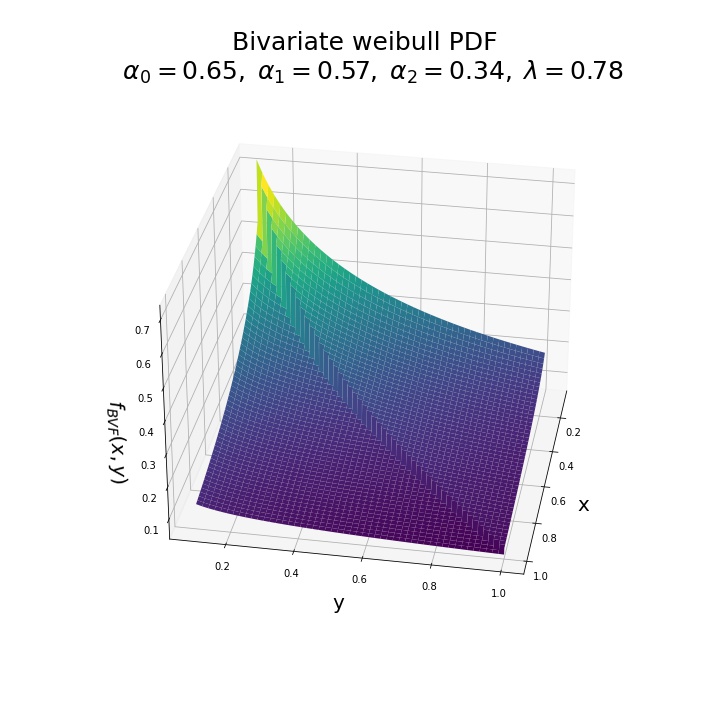}}
\subfloat{\includegraphics[width=.4\linewidth, height=.4\linewidth]{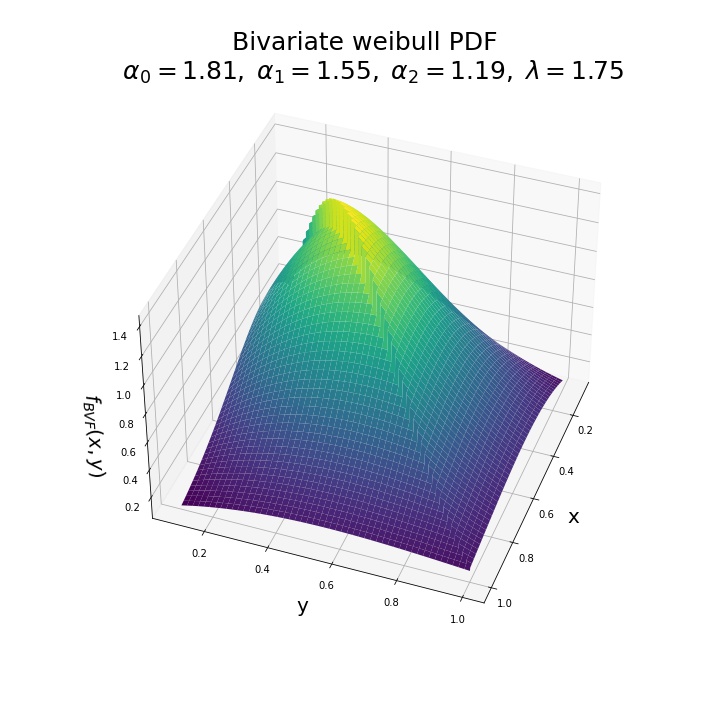}}
\hfill
\subfloat{\includegraphics[width=.4\linewidth, height=.4\linewidth]{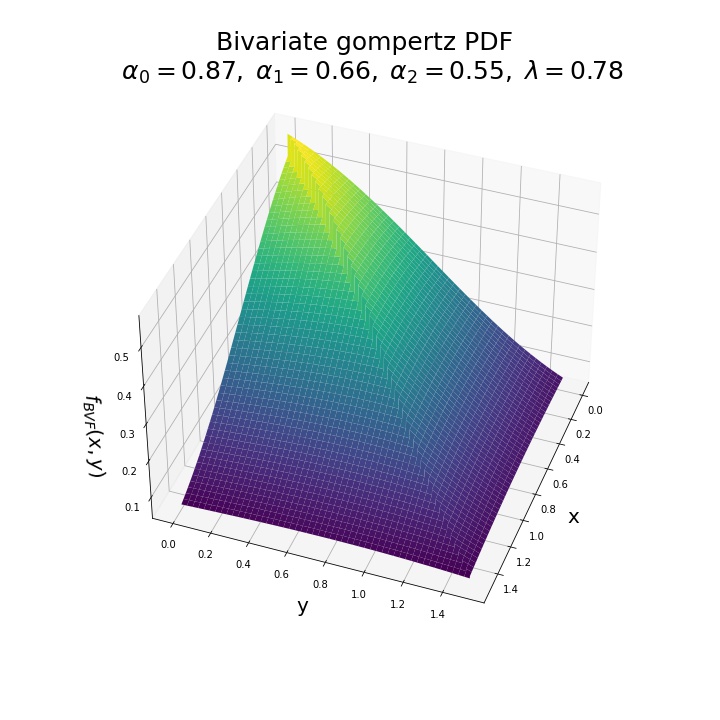}}
\subfloat{\includegraphics[width=.4\linewidth, height=.4\linewidth]{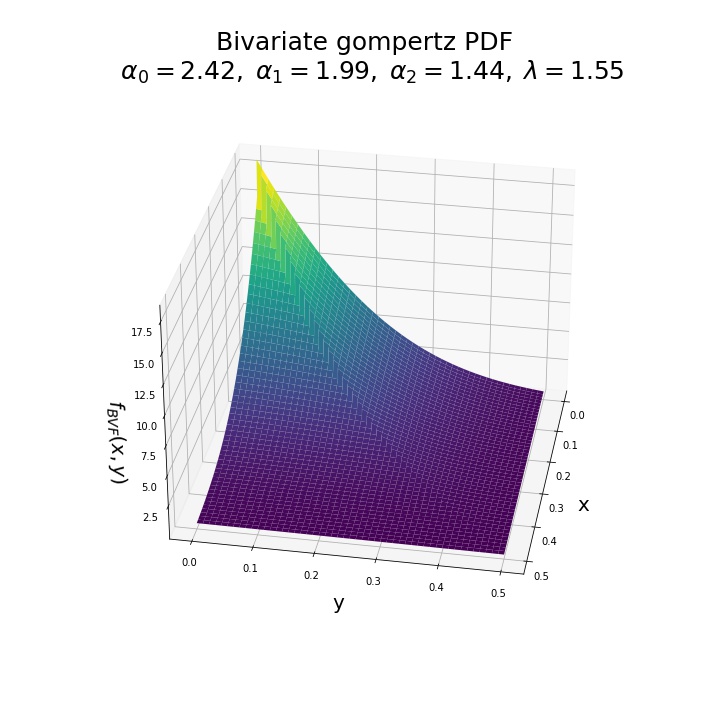}}
\hfill
\subfloat{\includegraphics[width=.4\linewidth, height=.4\linewidth]{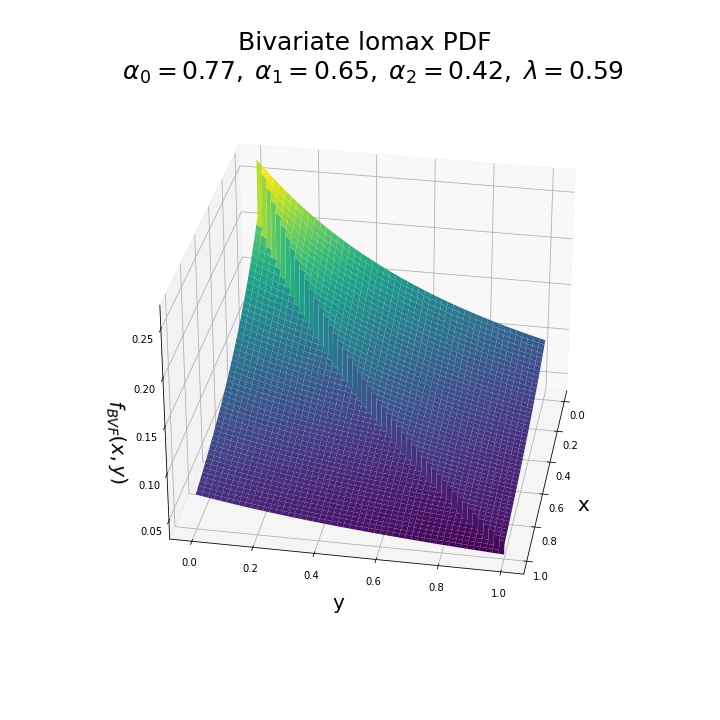}}
\subfloat{\includegraphics[width=.4\linewidth, height=.4\linewidth]{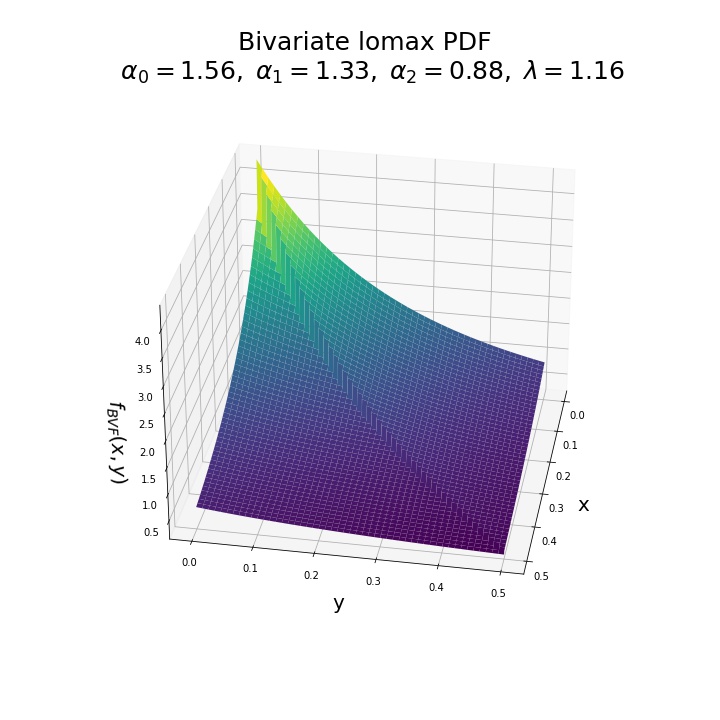}}
\caption{Surface plots of different members of the $BVF(\alpha_0, \alpha_1, \alpha_2, \lambda)$ distribution: bivariate Weibull, bivariate Gompertz, and bivariate Lomax (top to bottom).} \label{Surface}
\end{figure}

Let $U_0,\,U_1$, and $U_2$ denote three independent random variables, with the probability distribution of $U_i$ specified as 
\begin{equation}
   U_i \sim S(u_i;\lambda,\alpha_i) = \left( S_0(u_i;\lambda) \right)^{\alpha_i}, \quad u_i>0, \quad i=1,2,3. \label{shock}
\end{equation}
Define $X=\min\left\{ U_0,\,U_1 \right\}$ and
$Y=\min\left\{ U_0,\,U_2 \right\}$. Then, the joint survival function of $X$ and $Y$
is given by
\begin{eqnarray}
   & S_{X,\,Y}\left( x,\,y \right)
   &= P\left( \min\left\{ U_0,\,U_1 \right\}\geq x,\,\min\left\{ U_0,\,U_2
   \right\}\geq y \right) \nonumber \\
   &&= \begin{cases}
      S^{(1)}(x,\,y) & \text{if } 0<x<y<\infty\\
      S^{(2)}(x,\,y) & \text{if } 0<y<x<\infty\\
      S^{(3)}(x) & \text{if }0<x=y<\infty, \label{surv}
   \end{cases} 
\end{eqnarray}
where
\begin{align*}
   & S^{(1)}(x,\,y) = (S_0\left( y;\lambda \right))^{\alpha_0+\alpha_2}
   (S_0\left( x;\lambda \right))^{\alpha_1}, \\
   & S^{(2)}(x,\,y) = (S_0\left( x;\lambda \right))^{\alpha_0+\alpha_1}
   (S_0\left( y;\lambda \right))^{\alpha_2}, \\
   & S^{(3)}(x) = (S_0\left( x;\lambda \right))^{\alpha_0+\alpha_1+\alpha_2}.   
\end{align*}
From the joint survival function of $X$ and $Y$, their joint probability density function (JPDF) can be worked out. For $0<x<y<\infty$, the JPDF of $X$ and $Y$
is given by
\begin{align*}
   f^{(1)}(x,\,y) = \frac{\partial^2}{\partial x \partial y} S^{(1)}(x,\,y)
   = \alpha_1 \left( \alpha_0 + \alpha_2 \right) (S_0\left(
   y;\lambda \right))^{\alpha_0+\alpha_2-1} (S_0\left( x;\lambda \right))^{\alpha_1-1}f\left( x;\lambda \right)
   f\left( y;\lambda \right),
\end{align*}
where $f(\cdot;\,\lambda)$ is the probability density function
corresponding to the survival function $S_0(\cdot;\,\lambda)$. For
$0<y<x<\infty$, the JPDF of $X$ and $Y$ is
\begin{align*}
   f^{(2)}(x,\,y) = \frac{\partial^2}{\partial x \partial y} S^{(2)}(x,\,y)
   = \alpha_2 \left( \alpha_0 + \alpha_1 \right) (S_0\left( y;\lambda
   \right))^{\alpha_2-1} (S_0\left( x;\lambda \right))^{\alpha_0+\alpha_1-1}f\left( x;\lambda \right)
   f\left( y;\lambda \right).
\end{align*}
Therefore, we can calculate
\begin{align*}
   P\left( X<Y \right)
   &= \int_0^\infty \int_0^y \alpha_1 \left( \alpha_0 + \alpha_2 \right)
   S^{\alpha_0+\alpha_2-1}\left( y;\lambda \right) S^{\alpha_1-1}\left( x;\lambda
   \right)f\left( x;\lambda \right) f\left( y;\lambda \right)dxdy \\
   &= \int_0^\infty \left( \alpha_0+\alpha_2 \right) S^{\alpha_0+\alpha_2-1}\left(
   y;\lambda \right) f\left( y;\lambda \right) \left( 1 - S^{\alpha_1}\left(
   y;\lambda \right) \right) dy \\
   &= \frac{\alpha_1}{\alpha_0+\alpha_1+\alpha_2}.
\end{align*}
Similarly, we obtain
\begin{align*}
   P\left( Y<X \right) = \frac{\alpha_2}{\alpha_0+\alpha_1+\alpha_2}.
\end{align*}
From here, it is straightforward to see that
\begin{equation}
   P\left( X=Y \right) = 1 - P\left( X<Y \right) - P\left( Y<X \right) =
   \frac{\alpha_0}{\alpha_0+\alpha_1+\alpha_2}. \label{tie}
\end{equation}
This implies that the joint distribution of $X$ and $Y$ has a singular
component on the straight line $x=y$. The PDF of the singular part can be
obtained as follows:
\begin{align*}
   f^{(3)}\left( x \right) = P\left( X=Y \right) \left[ - \frac{\partial}{\partial x}
   S^{(3)}(x) \right] = \alpha_0 (S_0\left( x;\lambda
   \right))^{\alpha_0+\alpha_1+\alpha_2-1} f\left( x;\lambda \right).
\end{align*}
Summarizing the above, the JPDF of $X$ and $Y$ is given by
\begin{eqnarray}
   f_{X,\,Y}(x,\,y) 
   &= \begin{cases}
      f^{(1)}(x,\,y) & \text{if } 0<x<y<\infty\\
      f^{(2)}(x,\,y) & \text{if } 0<y<x<\infty\\
      f^{(3)}(x) & \text{if }0<x=y<\infty\\
      0 & \text{otherwise}. \label{JPDF}
   \end{cases} 
\end{eqnarray}
For convenience of reference, we say that $(X, Y)$ has a distribution in
bivariate Lehmann family if the joint survival function of $X$ and $Y$ is
given by Eq.\eqref{surv} or, equivalently, the JPDF is given by
Eq.\eqref{JPDF}, and we denote it as
\[
(X,Y) \sim BVF(\alpha_0, \alpha_1, \alpha_2, \lambda),
\]
for brevity. Clearly, by choosing different baseline survival functions in
the distributions of $U_i$s in Eq.\eqref{shock}, we arrive at different
bivariate models, for example bivariate Weibull, bivariate Gompertz, and
bivariate Lomax. Plots of the bivariate densities are presented in Figure
\ref{Surface}.

\section{Modelling dependent competing risks data} \label{sec:model-inference}
In this section, first, we describe the structure of dependent competing risks data we deal with in this paper. Then, likelihood inference for our proposed model, and various related issues are discussed. 
\subsection{Dependent competing risks data}
Consider two risk factors that can cause failure to each of the $n$ subjects enrolled in a survival study. The two risk factors are not assumed to be independent; that is, they can influence each other. Suppose $X$ and $Y$ are random variables denoting the lifetimes of a subject under the two risks factors, and the observed lifetime is $T = Min(X, Y)$. 

It is quite possible in reality that a subject fails from both the risk
factors simultaneously. For example, in complex medical studies, two risks
may activate simultaneously and are responsible for the event of interest
(e.g. death) to occur. Therefore, we consider a general set up where we
allow ties in the competing risks data.

The form of observed data is
$$Data = \left\{ (t_i, \delta_{i}),\, i=1,\,2,\,\ldots,\,n \right\},$$
where $t_i$ is the lifetime of the $i-$th unit, and $\delta_i$ is the indicator variable that gives information about the failure mode of the subjects, as follows: 
\begin{eqnarray}
   \delta_i =
   \begin{cases}
      0, & X_i = Y_i\\
      1, & X_i < Y_i \\ 
      2, & X_i > Y_i \\ 
      3, & X_i > C, Y_i > C, \label{indicator}
   \end{cases}
\end{eqnarray}
where $C$ is the right censoring time, usually at the end of a study period, implying a right censored lifetime when $\delta_i = 3$. 
For convenience of exposition, define index sets $A_k$ that contains
subjects with indicator variable $\delta = k$, , $k=0,\,1,\,2,\,3$,
\textit{i.e.},
\begin{align*}
   A_k = \left\{ i: \delta_i=k,\,i=1,\,2,\,\ldots,\,n \right\}.  
\end{align*}
Let $m_k = |A_k|$, $k=0,...,3$ be the cardinality (i.e., the number of observations) of the sets.  

\subsection{Maximum likelihood estimation}
The likelihood function is constructed considering contributions of subjects according to their failure status, as follows \\
(a) When $\delta=1$, the contribution is $-\frac{\partial}{\partial x}
S_{X,\,Y}(x,\,y)\bigg\vert_{x=t, y=t}$, \\
(b) When $\delta=2$, the contribution is $\frac{\partial}{\partial y} S_{X,\,Y}(x,\,y)\bigg\vert_{x=t, y=t}$, \\
(c) When $\delta=0$ (i.e., a tie between the risk factors), the
contribution is $f^{(3)}(t)$, and finally \\
(d) When $\delta=3$, the contribution is $S^{(3)}\left(t, t\right)$.

Combining these cases, the likelihood function without the multiplicative constant is given by
\begin{eqnarray}
   & L(\boldsymbol \theta) & \propto 
   \prod_{i \in A_1} \bigg[- \frac{\partial}{\partial x} S^{(1)}(x, y)\bigg\vert_{x=t_{i}, y=t_{i}}\bigg] \times \prod_{i \in A_2} \bigg[-\frac{\partial}{\partial y} S^{(2)}(x, y)\bigg\vert_{x=t_{i}, y=t_{i}}\bigg] \nonumber \\
   && \times \prod_{i \in A_0} \bigg[f^{(3)} \left( t_{i} \right)\bigg] \times \prod_{i \in A_3} \bigg[S^{(3)}\left( t_{i} \right)\bigg],
\end{eqnarray}
where $\boldsymbol \theta = (\alpha_0, \alpha_1, \alpha_2, \lambda)$. 
Then, the corresponding log-likelihood function, without the additive constant, is given by
\begin{align}
   \log L(\boldsymbol \theta) 
   &= \sum_{k=0}^{2} m_k \log\alpha_k + \left( \alpha_0+\alpha_1+\alpha_2 \right) \sum_{i=1}^{n} \log S\left(t_i;\lambda\right) + \sum_{i\in A_0\cup A_1\cup A_2} \log \frac{f\left( t_i;\lambda \right)}{S\left(t_i;\lambda\right)}. \label{eq:Loglik}
\end{align}

Equating the partial derivative of the log-likelihood function with respect to $\alpha_k$ to zero, we obtain
\begin{equation}
   \frac{m_k}{\alpha_k} + \sum_{i=1}^{n} \log S\left(t_i;\lambda\right) = 0 \Rightarrow \alpha_k = \frac{-m_k}{\sum_{i=1}^{n} \log S\left(t_i;\lambda\right)}, \quad k=0,1,2. \label{alphas}
\end{equation}
Substituting $\alpha_k$ in Eq.\eqref{eq:Loglik}, the profile log-likelihood in $\lambda$ is obtained as
\begin{align}
   p(\lambda)
   &= -(m_0 + m_1 + m_2) \log\bigg(-\sum_{i=1}^{n} \log S\left(t_i;\lambda\right)\bigg) + \sum_{i\in A_0\cup A_1\cup A_2} \log \frac{f\left( t_i;\lambda \right)}{S\left(t_i;\lambda\right)}.  \label{lambda}
\end{align}

The above derivations greatly simplifies the optimization of the log-likelihood function in this case, reducing it to a problem of a one-dimensional optimization. The profile log-likelihood in Eq.\eqref{lambda} can be maximized to get the MLE $\widehat{\lambda}$ of $\lambda$, which can then be plugged into Eq.\eqref{alphas} to obtain MLEs $\widehat{\alpha}_0$, $\widehat{\alpha}_1$, $\widehat{\alpha}_2$ of $\alpha_0, \alpha_1, \alpha_2$, respectively. For optimizing Eq.\eqref{lambda}, any routine one-dimensional optimizer from a standard statistical software may be used.     

\subsection{Asymptotic confidence intervals}
The asymptotic variance of the MLEs can be estimated by using the observed Fisher information matrix, which is the negative of the hessian of the log-likelihood function. That is, the observed Fisher information matrix $\boldsymbol I(\boldsymbol \theta)$ is
$$\boldsymbol I(\boldsymbol \theta) = -\nabla^2(\log L(\boldsymbol \theta)).$$
Due to asymptotic normality of MLEs, the distribution of
$\sqrt{n}(\widehat{\boldsymbol \theta} - \boldsymbol \theta)$ may be
approximated by a $N_4(\boldsymbol 0, \boldsymbol
I^{-1}(\widehat{\boldsymbol \theta}))$ distribution for large values of sample sizes.
Therefore, estimated asymptotic variances of $\widehat{\alpha}_0,
\widehat{\alpha}_1, \widehat{\alpha}_2$ and $\widehat{\lambda}$ are given
by the diagonal elements of $\boldsymbol I^{-1}(\widehat{\boldsymbol
\theta})$. Using these information, asymptotic 95\% confidence intervals
(CIs) for the parameters can be easily constructed; for example,
$$\alpha_0 \pm 1.96\sqrt{\widehat{Var({\widehat{\alpha}_0)}}},
$$
is an asymptotic 95\% CI for $\alpha_0$.    

Parametric bootstrap approach may be used for the confidence intervals as well. To implement an approach based on parametric bootstrap, we use the following algorithm: 

\vspace{2mm}
\noindent\textbf{Algorithm:}
\begin{enumerate}[label=\arabic*.,itemsep=-1ex]
   \item For a given data of size $n$, obtain MLE $\widehat\btheta$ of parameter $\btheta$ = $(\theta_1, \theta_2, \theta_3, \theta_4) = (\alpha_0, \alpha_1, \alpha_3, \lambda)$ 
   \item Using $\widehat\btheta$, generate a data of the same size $n$ from the assumed model 
   \item Based on the generated data, obtain MLE ${\widehat{\btheta}^*}$ 
   \item Repeat steps 2-3 $B$ times, to get $B$ bootstrap estimates
      ${\widehat\btheta^*_1, \, \widehat\btheta_2^*, \,\ldots,
      \,\widehat\btheta_B^*}$
   \item To construct bootstrap confidence interval for $\theta_i$, arrange
      $\widehat{\theta}^*_{i1},\,\ldots,\,\widehat\theta^*_{iB}$ (the $i$th
      components of ${\widehat\btheta^*_1, \, \widehat\btheta_2^*, \,\ldots,
      \,\widehat\btheta_B^*}$, respectively) in the
      ascending order to get $\widehat\theta^*_{i(1)} <
      \widehat\theta^*_{i(2)} < ... <\widehat\theta^*_{i(B)}$,
      $i=1,\,2,\,3,\,4$
   \item A $100(1-\alpha)\%$ parametric bootstrap confidence interval of $\theta_i$ is given by
      $\left(\widehat\theta^*_{i([B\frac{\alpha}{2}])},\,\widehat\theta^*_{i([B (1-\frac{\alpha}{2})])}\right)$.
\end{enumerate}

\subsection{Model Selection}
Marshall et al.~\cite{MMO} used a simple yet powerful approach to investigate whether data can correctly identify their parent distribution in the context of univariate distributions. Following the same, we use a model selection procedure which can be conveniently used for dependent competing risks data. 

The model selection procedure used is conceptually simple. Suppose
$\mathcal{M}_1$,...,$\mathcal{M}_l$ are $l$ candidate models for a
bivariate data, in a given context. The candidate models may or may not be
nested. Let the vector of parameters associated with these models be
${\boldsymbol \lambda}_1$, ..., ${\boldsymbol \lambda}_l$, respectively,
and that they all are of the same dimension; denote the MLEs of the
parameter vectors by ${\widehat{\boldsymbol \lambda}}_1$, ...,
${\widehat{\boldsymbol \lambda}}_l$, respectively. Then, for the given
bivariate data, the most suitable model among the candidate models is
$\mathcal{M}^*$ if
\begin{equation}
\widehat{L}(\mathcal{M^*}) =
Max\{\widehat{L}_{\mathcal{M}_1}(\widehat{\boldsymbol \lambda}_1),
\widehat{L}_{\mathcal{M}_2}(\widehat{\boldsymbol \lambda}_2), ...,
\widehat{L}_{\mathcal{M}_l}(\widehat{\boldsymbol \lambda}_l) \}, \label{model-sel}
\end{equation}
where $\widehat{L}_{\mathcal{M}_j}(\widehat{\boldsymbol \lambda}_j)$ is the maximized likelihood function evaluated at the MLE for the $j$-th candidate model, $j=1,...,l$, and $\widehat{L}(\mathcal{M^*})$ is the maximum of all those maximized likelihoods. 

When the parameter ${\boldsymbol \lambda}_1$, ..., ${\boldsymbol \lambda}_l$ are not of the same dimension, instead of using the values of maximized likelihoods for model selection, one may use Akaike's information criterion~\cite{Akaike} in Eq.\eqref{model-sel}. The Akaike's information criterion adjusts for the different number of parameters in a model by adding a penalty term to the maximized likelihood. 

The above procedure can be applied for model selection for dependent competing risks data as follows. Within the $BVF(\alpha_0, \alpha_1, \alpha_2, \lambda)$ family, there are three candidate models: bivariate Weibull, bivariate Gompertz, and bivariate Lomax distributions. For a given dependent competing risks data, we fit all the candidate models of the $BVF(\alpha_0, \alpha_1, \alpha_2, \lambda)$ family and select the one with the largest maximized likelihood value as the most suitable model for the data. Note that this procedure is expected to offer a final model that indeed is most suitable for a given dependent competing risks data, as the $BVF(\alpha_0, \alpha_1, \alpha_2, \lambda)$ family is very rich, containing bivariate Weibull, Gompertz, and Lomax as special cases.    

\section{Simulation Study} \label{sec:sim}
The goal of the simulation study conducted here is twofold. First, to
assess the performance of the MLEs and the confidence intervals for the
proposed model. Secondly, to examine the performance of the model selection
approach. We have observed through simulations that all the methods of
inference proposed here perform satisfactorily, evaluated by their
respective assessment criteria. The detailed results of the numerical
experiments are presented in this section. The simulation study is
performed by using the R software. 

For the simulation study, we first generate bivariate data from each of the
members of the $BVF(\alpha_0, \alpha_1, \alpha_2, \lambda)$ family
separately. After generating bivariate data $(X_i, Y_i)$ of a given size,
say $n$, it is converted to a competing risks data; \textit{i.e.}, a subject is a
failure from the first risk factor if $X_i < Y_i$, from the second risk
factor if $X_i > Y_i$, and the failure is a tie from both risk factors if
$X_i = Y_i$. Complete as well as right censored data are considered in the
simulations.        

\subsection{Performance of MLEs and confidence intervals} 
Three sample sizes are used: $n$ = 100, 200, and 400. Along with complete data, right censored data with roughly 20\% and 40\% censoring are also used. Tables \ref{table:wei} - \ref{table:lom} present the results of the numerical experiments. 

Performance of the MLEs are assessed by using relative mean squared error (MSE) and relative bias, as defined below for one of the parameters, say $\alpha_0$: 
\begin{equation}
\textrm{Relative MSE}(\alpha_0) = \frac{\textrm{MSE} (\widehat{\alpha}_0)}{(\alpha_0^*)^2}, \quad \textrm{Relative Bias}(\alpha_0) = \frac{\textrm{Bias}(\widehat{\alpha}_0)}{\alpha_0^*}.
\end{equation}
where the true value of $\alpha_0$ is $\alpha_0^*$. For the confidence intervals, the coverage probabilities are estimated by the Monte Carlo probabilities of including the true parameter value.   

We observe that the MLEs, for all the models, have quite reasonable bias and MSE. As expected, the bias and MSE reduce with increasing sample size. Also, censoring has adverse effects on bias and MSE of MLEs for all the models: the higher the censoring, the more the MLEs suffer. Relatively, inference for bivariate Gompertz distribution based on dependent competing risks data seems to be the most affected due to censoring. Inference for bivariate Lomax is also affected to some extent. However, for censored data, the models are less affected for larger sample sizes compared to smaller sample sizes. Censoring seems to have the least effect on inference for the bivariate Weibull distribution among the three models. 

The confidence intervals also render satisfactory performance with respect to average length and coverage probability. The coverage probabilities for both types of confidence intervals are quite close to the nominal confidence level of 95\%. And it is also noteworthy that even with an increase in sample size, the confidence intervals can retain their coverage probability in spite of their average lengths reduce. The confidence intervals from the parametric bootstrap approach, in a relative sense, has a slightly lower coverage probability than the intervals calculated using the observed Fisher information matrix. 

In summary, we can conclude that the proposed method to obtain MLEs and confidence intervals for the model parameters of the $BVF(\alpha_0, \alpha_1, \alpha_2, \lambda)$ family of distributions based on dependent competing risks data performs very well.

\begin{table}[htb]\scriptsize
	\caption{Performance of the MLEs and the CIs of the parameters of bivariate Weibull ($\alpha_0 = 1.34$, $\alpha_1 = 1.17$, $\alpha_2 = 0.86$, $\lambda = 0.91$) distribution based on dependent competing risks data.}
	\begin{center}
		\begin{tabular}{| c | c | c | c | c | c | c | c |}
			\toprule
			\multicolumn{8}{| c |}{Complete Data} \\
			\midrule
			&  & \multicolumn{2}{c|}{Point Estimate} &    	
         	\multicolumn{2}{c|}{\makecell{95\% CI \\ (Asymptotic)}} &
         	\multicolumn{2}{c|}{\makecell{95\% CI \\ (Bootstrapping)}} \\
			\cmidrule(lr){3-4}\cmidrule(lr){5-6}\cmidrule(lr){7-8}
			Sample& Parameters & Relative & Relative & Average & Coverage & Average& Coverage\\
			Size & & MSE & Bias & Length & Percentage & Length & Percentage \\ 
			\midrule
			\multirow{4}{*}{100}
			& $\alpha_0$ & 0.031 & 0.028 & 0.930 & 0.954 & 0.788 & 0.894 \\
			& $\alpha_1$ & 0.039 & 0.041 & 0.867 & 0.951 & 0.744 & 0.876 \\
			& $\alpha_2$ & 0.049 & 0.044 & 0.730 & 0.949 & 0.622 & 0.892 \\
			& $\lambda$ & 0.007 & 0.018 & 0.284 & 0.934 & 0.241 & 0.874 \\
			\midrule
			\multirow{4}{*}{200} 
			& $\alpha_0$ & 0.014 & 0.012 & 0.641 & 0.955 & 0.545 & 0.880 \\
			& $\alpha_1$ & 0.016 & 0.013 & 0.594 & 0.958 & 0.500 & 0.906 \\
			& $\alpha_2$ & 0.023 & 0.012 & 0.501 & 0.940 & 0.421 & 0.884 \\
			& $\lambda$ & 0.003 & 0.006 & 0.198 & 0.940 & 0.165 & 0.896 \\
			\midrule
			\multirow{4}{*}{400} 
			& $\alpha_0$ & 0.007 & 0.004 & 0.449 & 0.953 & 0.374 & 0.894 \\
			& $\alpha_1$ & 0.008 & 0.003 & 0.416 & 0.945 & 0.347 & 0.902 \\
			& $\alpha_2$ & 0.011 & 0.010 & 0.352 & 0.944 & 0.294 & 0.884 \\
			& $\lambda$ & 0.001 & 0.004 & 0.139 & 0.950 & 0.116 & 0.902 \\
			\bottomrule
		\end{tabular}
		\begin{tabular}{| c | c | c | c | c | c | c | c |}
		\toprule
		\multicolumn{8}{| c |}{Data with 20\% censoring} \\
			\midrule
			&  & \multicolumn{2}{c|}{Point Estimate} &    	
         	\multicolumn{2}{c|}{\makecell{95\% CI \\ (Asymptotic)}} &
         	\multicolumn{2}{c|}{\makecell{95\% CI \\ (Bootstrapping)}} \\
			\cmidrule(lr){3-4}\cmidrule(lr){5-6}\cmidrule(lr){7-8}
			Sample& Parameters & Relative & Relative & Average & Coverage & Average& Coverage\\
			Size & & MSE & Bias & Length & Percentage & Length & Percentage \\ 
			\midrule
			\multirow{4}{*}{100}
			& $\alpha_0$ & 0.046 & 0.031 & 1.065 & 0.944 & 0.912 & 0.900 \\
			& $\alpha_1$ & 0.047 & 0.021 & 0.973 & 0.954 & 0.844 & 0.898 \\
			& $\alpha_2$ & 0.060 & 0.027 & 0.811 & 0.947 & 0.700 & 0.898 \\
			& $\lambda$ & 0.009 & 0.014 & 0.329 & 0.945 & 0.278 & 0.878 \\
			\midrule
			\multirow{4}{*}{200} 
			& $\alpha_0$ & 0.021 & 0.015 & 0.742 & 0.951 & 0.625 & 0.896 \\
			& $\alpha_1$ & 0.023 & 0.009 & 0.679 & 0.929 & 0.569 & 0.874 \\
			& $\alpha_2$ & 0.030 & 0.018 & 0.567 & 0.941 & 0.474 & 0.910 \\
			& $\lambda$ & 0.004 & 0.007 & 0.231 & 0.948 & 0.193 & 0.876 \\
			\midrule
			\multirow{4}{*}{400} 
			& $\alpha_0$ & 0.009 & 0.006 & 0.519 & 0.956 & 0.432 & 0.882 \\
			& $\alpha_1$ & 0.010 & 0.007 & 0.478 & 0.954 & 0.397 & 0.892 \\
			& $\alpha_2$ & 0.013 & 0.000 & 0.395 & 0.945 & 0.326 & 0.918 \\
			& $\lambda$ & 0.002 & 0.003 & 0.163 & 0.942 & 0.135 & 0.902 \\
			\bottomrule
		\end{tabular}
		\begin{tabular}{| c | c | c | c | c | c | c | c |}
		\toprule
		\multicolumn{8}{| c |}{Data with 40\% censoring} \\
			\midrule
			&  & \multicolumn{2}{c|}{Point Estimate} &    	
         	\multicolumn{2}{c|}{\makecell{95\% CI \\ (Asymptotic)}} &
         	\multicolumn{2}{c|}{\makecell{95\% CI \\ (Bootstrapping)}} \\
			\cmidrule(lr){3-4}\cmidrule(lr){5-6}\cmidrule(lr){7-8}
			Sample& Parameters & Relative & Relative & Average & Coverage & Average& Coverage\\
			Size & & MSE & Bias & Length & Percentage & Length & Percentage \\ 
			\midrule
			\multirow{4}{*}{100}
			& $\alpha_0$ & 0.063 & 0.039 & 1.342 & 0.948 & 1.158 & 0.898 \\
			& $\alpha_1$ & 0.077 & 0.038 & 1.221 & 0.942 & 1.049 & 0.888 \\
			& $\alpha_2$ & 0.080 & 0.019 & 0.982 & 0.943 & 0.849 & 0.872 \\
			& $\lambda$ & 0.011 & 0.012 & 0.391 & 0.956 & 0.328 & 0.894 \\
			\midrule
			\multirow{4}{*}{200} 
			& $\alpha_0$ & 0.033 & 0.024 & 0.932 & 0.951 & 0.786 & 0.894 \\
			& $\alpha_1$ & 0.035 & 0.017 & 0.846 & 0.945 & 0.714 & 0.878 \\
			& $\alpha_2$ & 0.042 & 0.021 & 0.692 & 0.946 & 0.582 & 0.898 \\
			& $\lambda$ & 0.005 & 0.006 & 0.275 & 0.952 & 0.230 & 0.900 \\
			\midrule
			\multirow{4}{*}{400} 
			& $\alpha_0$ & 0.015 & 0.008 & 0.649 & 0.952 & 0.545 & 0.900 \\
			& $\alpha_1$ & 0.017 & 0.007 & 0.591 & 0.942 & 0.479 & 0.902 \\
			& $\alpha_2$ & 0.021 & 0.003 & 0.481 & 0.945 & 0.405 & 0.890 \\
			& $\lambda$ & 0.002 & 0.003 & 0.194 & 0.956 & 0.161 & 0.898 \\
			\bottomrule
		\end{tabular}
		\label{table:wei}
	\end{center}
\end{table}

\begin{table}[htb]\scriptsize
	\caption{Performance of the MLEs and the CIs of the parameters of bivariate Gompertz ($\alpha_0 = 1.13$, $\alpha_1 = 0.96$, $\alpha_2 = 0.79$, $\lambda = 1.05$) distribution based on dependent competing risks data.}
	\begin{center}
		\begin{tabular}{| c | c | c | c | c | c | c | c |}
			\toprule
			\multicolumn{8}{| c |}{Complete data} \\
			\midrule
			&  & \multicolumn{2}{c|}{Point Estimate} &    	
         	\multicolumn{2}{c|}{\makecell{95\% CI \\ (Asymptotic)}} &
         	\multicolumn{2}{c|}{\makecell{95\% CI \\ (Bootstrapping)}} \\
			\cmidrule(lr){3-4}\cmidrule(lr){5-6}\cmidrule(lr){7-8}
			Sample& Parameters & Relative & Relative & Average & Coverage & Average& Coverage\\
			Size & & MSE & Bias & Length & Percentage & Length & Percentage \\ 
			\midrule
			\multirow{4}{*}{100}
			& $\alpha_0$ & 0.227 & -0.029 & 2.800 & 0.845 & 2.110 & 0.854 \\
			& $\alpha_1$ & 0.212 & -0.042 & 2.352 & 0.833 & 1.827 & 0.854 \\
			& $\alpha_2$ & 0.230 & -0.040 & 1.959 & 0.838 & 1.478 & 0.862 \\
			& $\lambda$ & 0.211 & 0.182 & 1.926 & 0.955 & 1.563 & 0.824 \\
			\midrule
			\multirow{4}{*}{200} 
			& $\alpha_0$ & 0.170 & 0.023 & 2.089 & 0.895 & 1.686 & 0.902 \\
			& $\alpha_1$ & 0.192 & 0.037 & 1.814 & 0.886 & 1.448 & 0.906 \\
			& $\alpha_2$ & 0.187 & 0.030 & 1.492 & 0.880 & 1.218 & 0.868 \\
			& $\lambda$ & 0.100 & 0.074 & 1.333 & 0.963 & 1.093 & 0.884 \\
			\midrule
			\multirow{4}{*}{400} 
			& $\alpha_0$ & 0.091 & 0.023 & 1.416 & 0.914 & 1.185 & 0.884 \\
			& $\alpha_1$ & 0.092 & 0.017 & 1.204 & 0.914 & 1.013 & 0.890 \\
			& $\alpha_2$ & 0.093 & 0.022 & 1.003 & 0.909 & 0.836 & 0.892 \\
			& $\lambda$ & 0.047 & 0.033 & 0.932 & 0.952 & 0.776 & 0.886 \\
			\bottomrule
		\end{tabular}
		\begin{tabular}{| c | c | c | c | c | c | c | c |}
		\toprule
		\multicolumn{8}{| c |}{Data with 20\% censoring} \\
			\midrule
			&  & \multicolumn{2}{c|}{Point Estimate} &    	
         	\multicolumn{2}{c|}{\makecell{95\% CI \\ (Asymptotic)}} &
         	\multicolumn{2}{c|}{\makecell{95\% CI \\ (Bootstrapping)}} \\
			\cmidrule(lr){3-4}\cmidrule(lr){5-6}\cmidrule(lr){7-8}
			Sample& Parameters & Relative & Relative & Average & Coverage & Average& Coverage\\
			Size & & MSE & Bias & Length & Percentage & Length & Percentage \\ 
			\midrule
			\multirow{4}{*}{100}
			& $\alpha_0$ & 2.047 & 0.323 & 11.67 & 0.817 & 3.618 & 0.884 \\
			& $\alpha_1$ & 2.338 & 0.354 & 10.32 & 0.809 & 3.180 & 0.882 \\
			& $\alpha_2$ & 2.095 & 0.327 & 8.154 & 0.813 & 2.589 & 0.900 \\
			& $\lambda$ & 0.439 & 0.199 & 3.119 & 0.972 & 2.098 & 0.876 \\
			\midrule
			\multirow{4}{*}{200} 
			& $\alpha_0$ & 1.376 & 0.273 & 6.945 & 0.872 & 3.406 & 0.918 \\
			& $\alpha_1$ & 1.533 & 0.288 & 6.074 & 0.873 & 2.965 & 0.930 \\
			& $\alpha_2$ & 1.456 & 0.266 & 4.891 & 0.866 & 2.383 & 0.918 \\
			& $\lambda$ & 0.233 & 0.090 & 2.206 & 0.969 & 1.578 & 0.928 \\
			\midrule
			\multirow{4}{*}{400} 
			& $\alpha_0$ & 0.901 & 0.248 & 4.140 & 0.899 & 3.120 & 0.904 \\
			& $\alpha_1$ & 0.874 & 0.239 & 3.488 & 0.902 & 2.619 & 0.908 \\
			& $\alpha_2$ & 0.930 & 0.251 & 2.921 & 0.897 & 2.192 & 0.886 \\
			& $\lambda$ & 0.133 & 0.007 & 1.562 & 0.960 & 1.173 & 0.902 \\
			\bottomrule
		\end{tabular}
		\begin{tabular}{| c | c | c | c | c | c | c | c |}
		\toprule
		\multicolumn{8}{| c |}{Data with 40\% censoring} \\
			\midrule
			&  & \multicolumn{2}{c|}{Point Estimate} &    	
         	\multicolumn{2}{c|}{\makecell{95\% CI \\ (Asymptotic)}} &
         	\multicolumn{2}{c|}{\makecell{95\% CI \\ (Bootstrapping)}} \\
			\cmidrule(lr){3-4}\cmidrule(lr){5-6}\cmidrule(lr){7-8}
			Sample& Parameters & Relative & Relative & Average & Coverage & Average& Coverage\\
			Size & & MSE & Bias & Length & Percentage & Length & Percentage \\ 
			\midrule
			\multirow{4}{*}{100}
			& $\alpha_0$ & 2.123 & 0.178 & 20.30 & 0.731 & 2.791 & 0.814 \\
			& $\alpha_1$ & 2.139 & 0.172 & 17.09 & 0.730 & 2.640 & 0.810 \\
			& $\alpha_2$ & 2.254 & 0.189 & 14.32 & 0.732 & 2.152 & 0.798 \\
			& $\lambda$ & 1.430 & 0.654 & 5.673 & 0.964 & 3.657 & 0.802 \\
			\midrule
			\multirow{4}{*}{200} 
			& $\alpha_0$ & 1.924 & 0.266 & 14.00 & 0.821 & 3.646 & 0.870 \\
			& $\alpha_1$ & 2.077 & 0.260 & 12.04 & 0.808 & 3.168 & 0.874 \\
			& $\alpha_2$ & 2.045 & 0.255 & 9.845 & 0.806 & 2.566 & 0.868 \\
			& $\lambda$ & 0.648 & 0.315 & 4.014 & 0.969 & 2.596 & 0.870 \\
			\midrule
			\multirow{4}{*}{400} 
			& $\alpha_0$ & 1.486 & 0.253 & 8.810 & 0.843 & 3.600 & 0.918 \\
			& $\alpha_1$ & 1.403 & 0.247 & 7.367 & 0.833 & 3.049 & 0.918 \\
			& $\alpha_2$ & 1.391 & 0.241 & 6.031 & 0.839 & 2.547 & 0.918 \\
			& $\lambda$ & 0.353 & 0.169 & 2.844 & 0.968 & 1.973 & 0.922 \\
			\bottomrule
		\end{tabular}
		\label{table:gom}
	\end{center}
\end{table}

\begin{table}[htb]\scriptsize
	\caption{Performance of the MLEs and the CIs of the parameters of bivariate Lomax ($\alpha_0 = 0.85$, $\alpha_1 = 0.57$, $\alpha_2 = 0.74$, $\lambda = 0.69$) distribution based on dependent competing risks data.}
	\begin{center}
		\begin{tabular}{| c | c | c | c | c | c | c | c |}
		\toprule
		\multicolumn{8}{| c |}{Complete data} \\
			\midrule
			&  & \multicolumn{2}{c|}{Point Estimate} &    	
         	\multicolumn{2}{c|}{\makecell{95\% CI \\ (Asymptotic)}} &
         	\multicolumn{2}{c|}{\makecell{95\% CI \\ (Bootstrapping)}} \\
			\cmidrule(lr){3-4}\cmidrule(lr){5-6}\cmidrule(lr){7-8}
			Sample& Parameters & Relative & Relative & Average & Coverage & Average& Coverage\\
			Size & & MSE & Bias & Length & Percentage & Length & Percentage \\ 
			\midrule
			\multirow{4}{*}{100}
			& $\alpha_0$ & 0.793 & 0.222 & 2.034 & 0.945 & 2.238 & 0.884 \\
			& $\alpha_1$ & 0.742 & 0.210 & 1.383 & 0.952 & 1.544 & 0.872 \\
			& $\alpha_2$ & 0.743 & 0.214 & 1.754 & 0.945 & 1.965 & 0.886 \\
			& $\lambda$ & 0.218 & 0.174 & 1.234 & 0.912 & 1.010 & 0.894 \\
			\midrule
			\multirow{4}{*}{200} 
			& $\alpha_0$ & 0.129 & 0.096 & 0.990 & 0.968 & 0.983 & 0.860 \\
			& $\alpha_1$ & 0.132 & 0.102 & 0.694 & 0.957 & 0.680 & 0.892 \\
			& $\alpha_2$ & 0.120 & 0.094 & 0.869 & 0.964 & 0.871 & 0.884 \\
			& $\lambda$ & 0.093 & -0.014 & 0.843 & 0.926 & 0.705 & 0.862 \\
			\midrule
			\multirow{4}{*}{400} 
			& $\alpha_0$ & 0.039 & 0.045 & 0.622 & 0.956 & 0.580 & 0.904 \\
			& $\alpha_1$ & 0.048 & 0.047 & 0.438 & 0.955 & 0.405 & 0.872 \\
			& $\alpha_2$ & 0.042 & 0.044 & 0.549 & 0.963 & 0.515 & 0.888 \\
			& $\lambda$ & 0.052 & -0.005 & 0.594 & 0.930 & 0.488 & 0.870 \\
			\bottomrule
		\end{tabular}
		\begin{tabular}{| c | c | c | c | c | c | c | c |}
		\toprule
		\multicolumn{8}{| c |}{Data with 20\% censoring} \\
			\midrule
			&  & \multicolumn{2}{c|}{Point Estimate} &    	
         	\multicolumn{2}{c|}{\makecell{95\% CI \\ (Asymptotic)}} &
         	\multicolumn{2}{c|}{\makecell{95\% CI \\ (Bootstrapping)}} \\
			\cmidrule(lr){3-4}\cmidrule(lr){5-6}\cmidrule(lr){7-8}
			Sample& Parameters & Relative & Relative & Average & Coverage & Average& Coverage\\
			Size & & MSE & Bias & Length & Percentage & Length & Percentage \\ 
			\midrule
			\multirow{4}{*}{100}
			& $\alpha_0$ & 9.741 & 0.417 & 5.996 & 0.894 & 2.319 & 0.940 \\
			& $\alpha_1$ & 10.84 & 0.434 & 4.107 & 0.889 & 1.601 & 0.944 \\
			& $\alpha_2$ & 15.86 & 0.459 & 5.287 & 0.903 & 1.991 & 0.930 \\
			& $\lambda$ & 0.579 & 0.208 & 2.170 & 0.976 & 1.829 & 0.928 \\
			\midrule
			\multirow{4}{*}{200}
			& $\alpha_0$ & 0.900 & 0.186 & 2.361 & 0.915 & 1.752 & 0.924 \\
			& $\alpha_1$ & 0.790 & 0.186 & 1.603 & 0.914 & 1.191 & 0.920 \\
			& $\alpha_2$ & 0.862 & 0.179 & 2.047 & 0.914 & 1.500 & 0.916 \\
			& $\lambda$ & 0.276 & 0.073 & 1.442 & 0.962 & 1.198 & 0.920 \\
			\midrule
			\multirow{4}{*}{400}
			& $\alpha_0$ & 0.128 & 0.063 & 1.166 & 0.915 & 1.121 & 0.896 \\
			& $\alpha_1$ & 0.139 & 0.072 & 0.804 & 0.939 & 0.761 & 0.878 \\
			& $\alpha_2$ & 0.132 & 0.068 & 1.026 & 0.930 & 0.979 & 0.904 \\
			& $\lambda$ & 0.127 & 0.041 & 0.996 & 0.965 & 0.832 & 0.902 \\
			\bottomrule
		\end{tabular}
		\begin{tabular}{| c | c | c | c | c | c | c | c |}
		\toprule
		\multicolumn{8}{| c |}{Data with 40\% censoring} \\
			\midrule
			&  & \multicolumn{2}{c|}{Point Estimate} &    	
         	\multicolumn{2}{c|}{\makecell{95\% CI \\ (Asymptotic)}} &
         	\multicolumn{2}{c|}{\makecell{95\% CI \\ (Bootstrapping)}} \\
			\cmidrule(lr){3-4}\cmidrule(lr){5-6}\cmidrule(lr){7-8}
			Sample& Parameters & Relative & Relative & Average & Coverage & Average& Coverage\\
			Size & & MSE & Bias & Length & Percentage & Length & Percentage \\ 
			\midrule
			\multirow{4}{*}{100}
			& $\alpha_0$ & 9.929 & 0.387 & 9.482 & 0.820 & 2.513 & 0.910 \\
			& $\alpha_1$ & 13.02 & 0.400 & 6.287 & 0.837 & 1.751 & 0.912 \\
			& $\alpha_2$ & 11.17 & 0.410 & 8.461 & 0.829 & 2.263 & 0.902 \\
			& $\lambda$ & 1.852 & 0.548 & 3.445 & 0.986 & 2.996 & 0.912 \\
			\midrule
			\multirow{4}{*}{200} 
			& $\alpha_0$ & 2.470 & 0.238 & 5.212 & 0.881 & 2.292 & 0.938 \\
			& $\alpha_1$ & 2.673 & 0.241 & 3.580 & 0.877 & 1.545 & 0.928 \\
			& $\alpha_2$ & 2.483 & 0.233 & 4.538 & 0.873 & 2.005 & 0.934 \\
			& $\lambda$ & 0.599 & 0.258 & 2.203 & 0.983 & 1.788 & 0.932 \\
			\midrule
			\multirow{4}{*}{400}
			& $\alpha_0$ & 0.309 & 0.092 & 2.979 & 0.901 & 1.851 & 0.930\\
			& $\alpha_1$ & 0.317 & 0.091 & 1.403 & 0.902 & 1.247 & 0.928 \\
			& $\alpha_2$ & 0.339 & 0.089 & 1.825 & 0.893 & 1.625 & 0.926 \\
			& $\lambda$ & 0.261 & 0.119 & 1.482 & 0.982 & 1.172 & 0.932 \\
			\bottomrule
		\end{tabular}
		\label{table:lom}
	\end{center}
\end{table}

\subsection{A study of model selection}
The simulation study on model selection for dependent competing risks data
is carried out within the bivariate family $BVF(\alpha_0, \alpha_1,
\alpha_2, \lambda)$. Here, we consider three models, \textit{viz.},
bivariate Weibull, bivariate Gompertz, and bivariate Lomax distributions
for comparison. The general approach is the following: we generate
dependent competing risks data from a parent distribution belonging to the
family $BVF(\alpha_0, \alpha_1, \alpha_2, \lambda)$, and then fit the
candidate models to the generated data. The set of candidate models
includes the parent distribution itself, along with other model(s) from the
family. We use both two- and three-model set ups in the study. This process
of generating data and fitting the candidate models to it is repeated a
large number of times, and the proportions of times each of the candidate
models are selected as the best model are recorded. For each of the
candidate models, this proportion is the Monte Carlo probability of being
selected as the model of choice for a given dependent competing risks data. 

\begin{figure}[t]
\centering
\subfloat{\includegraphics[width=0.9\linewidth, height=1.0\linewidth]{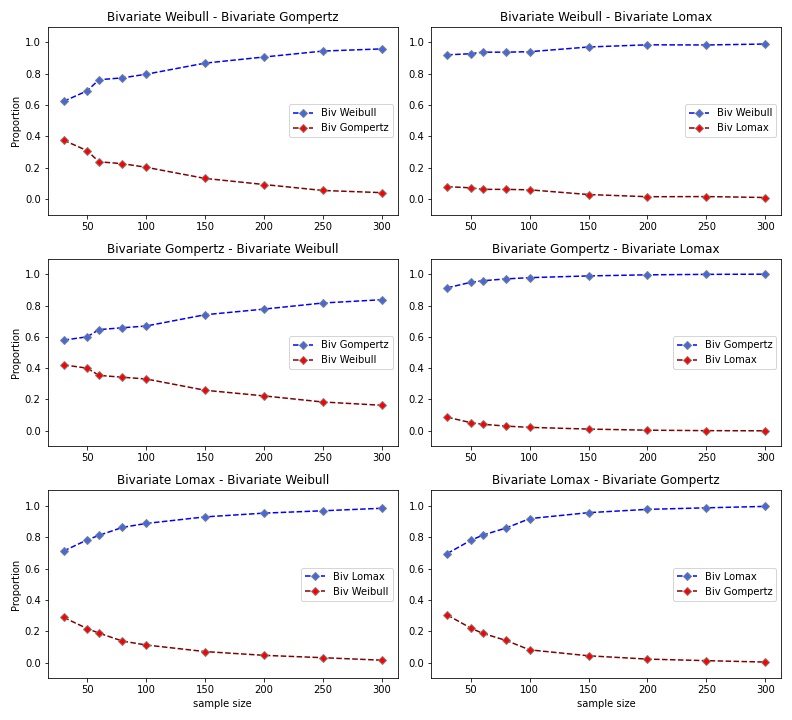}}
\caption{Empirical probability of model selection in the 2-model setting; for each plot, the ``AA-BB'' caption above the plot indicates AA as the parent distribution, and BB as the model fitted to the data other than the parent.} \label{fig:2-model}
\end{figure}

\begin{figure}[t]
\centering
\subfloat{\includegraphics[width=0.6\linewidth, height=1.0\linewidth]{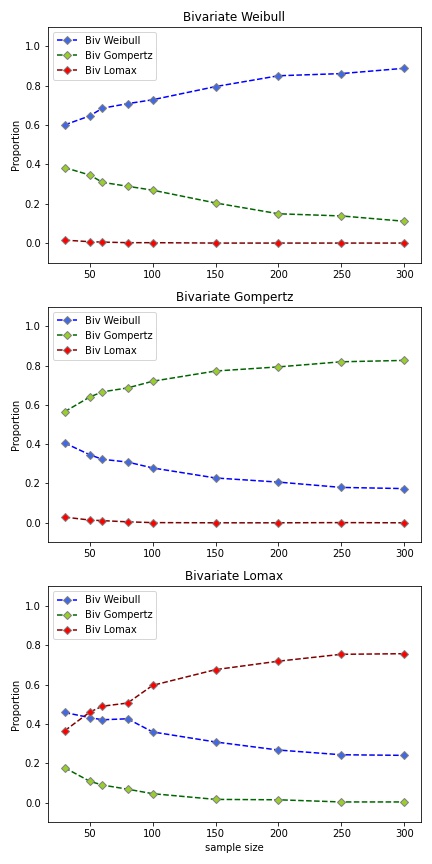}}
\caption{Empirical probability of model selection in the 3-model setting; the parent distributions for generating the data are indicated in captions above each plot.} \label{fig:3-model}
\end{figure}

There are two motivating factors for the study on model selection: first, the direct goal is to see whether the likelihood-based approach of Marshall et al.\cite{MMO}, used for univariate models, is successful in identifying the parent distribution in case of dependent competing risks data as well. In case it is successful, this approach can then be used as an effective tool for model selection in case of parametric modelling of dependent competing risks data. Secondly, as a tangential topic, this study will also indicate the relative richness of the bivariate distributions of the members of the family $BVF(\alpha_0, \alpha_1, \alpha_2, \lambda)$, in the sense of accommodating dependent competing risks data. Figures \ref{fig:2-model} and \ref{fig:3-model} give the results of the simulation study on model selection with the two-model and three-model set ups, respectively.

\subsubsection{Two-model comparisons based on dependent competing risks data}
In a two-model setting, it is clear that the parent distribution, whatever it is within the bivariate family, will be selected as the model of choice when the sample size is large. For small to moderate sample sizes, however, there are some interesting observations. Clearly, when the parent is bivariate Weibull, bivariate Gompertz is a weak choice with its probability of being selected dropping below 20\% when the sample size exceeds 100. However, for bivariate Gompertz as the parent, bivariate Weibull has a significant chance of being selected; even for sample size 150, the probability is above 20\%. Bivariate Lomax is always a very weak choice when data are generated from either bivariate Weibull or Gompertz. Similarly, when bivariate Lomax is the parent, selection probabilities for both bivariate Weibull and Gompertz drop rapidly with increasing sample size. 

From the above observations on the two-model comparisons within the family $BVF(\alpha_0, \alpha_1, \alpha_2, \lambda)$ based on dependent competing risks data, we can conclude that the likelihood-based model selection approach can identify the parent model correctly, with increasing confidence as the sample size increases. Moreover, we can also conclude that for modelling dependent competing risks data, the bivariate Weibull and the bivariate Gompertz distributions seem to be far apart in nature compared to the bivariate Lomax distribution, although all of them belong to the same family. Finally, we can also perhaps conclude that between bivariate Weibull and Gompertz, the former seems to be stronger as it fits well to dependent competing risks data generated from the latter, especially for small to moderate sample sizes. 

\subsubsection{Three-model comparisons based on dependent competing risks data}
In the three-model setting, the bivariate Lomax has almost no chance of
being selected when the parent is either bivariate Weibull or bivariate Gompertz. In contrast, bivariate Weibull and Gompertz give a fair amount of competition to each other: each of them having a significant probability of being selected when the other is the parent. It is also clear that bivariate Weibull is a stronger model than the bivariate Gompertz, with the former's probability of being selected ($\sim$20\%) as double the latter's probability of being selected ($\sim$10\%) even when the sample is large (300).  Moreover, when the parent is bivariate Lomax, bivariate Weibull remains a strong choice throughout, with its selection probability about 25\% even for large sample size. Nonetheless, it is also clear that the likelihood-based approach eventually picks the parent distribution of the dependent competing risks data correctly, whatever is the parent distribution, as sample size increases. 

In summary, it is quite clear that for dependent competing risks data, the likelihood-based approach is successful in identifying the parent distribution correctly, in both two-model and three-model settings. Also, we can conclude that the bivariate Weibull is the strongest model in this family $BVF(\alpha_0, \alpha_1, \alpha_2, \lambda)$ to model this data structure, as we have observed that apart from being the best choice for data generated from itself, for data generated from the other two parents as well, bivariate Weibull is a good choice.

In view of the above discussions, it is clear that likelihood-based model selection approach is quite successful for dependent competing risks data. That is, for a given competing risks data with the structure as dealt with in this paper, the likelihood-based approach as described above can be used to determine the appropriate model.   

\section{Analysis of real data} \label{sec:real-data}
The Diabetic Retinopathy Study conducted by the National Eye Institute,
United Sates of America was a clinical trial to evaluate laser-based treatment for patients with proliferative diabetic retinopathy. In this study, a total of 1758 patients were enrolled during the period 1972 to 1975. At enrollment, the minimum best corrected visual acuity in each eye was 20/100 for each of the study subjects. For each subject, one of the eyes was given a laser-based treatment, and the other eye was left untreated. Cs\"{o}rg\"o and Welsh~\cite{Csorgo} reported the uncensored part of the data for white male patients who received argon laser treatment which was one of three types of laser-based treatment given to the enrolled patients. For the data reported in Cs\"{o}rg\"o and Welsh~\cite{Csorgo}, failure of an eye, measured in days, was defined as the first time the best corrected visual acuity was below 5/200. For more details regarding the data, refer to Cs\"{o}rg\"o and Welsh~\cite{Csorgo}.

The diabetic retinopathy data can be looked at as a competing risks data, as analysed by Feizjavadian and Hashemi~\cite{Feiz2015} in the following way. Let $X_i$ and $Y_i$ denote the time to failure for the treated and the untreated eye, respectively for the $i$-th patient. Define $T_i$ = $Min(X_i, Y_i)$ as the time to blindness. Also, it is easy to define an indicator variable $\nu_i$ such that 
\begin{eqnarray}
   \nu_i =
   \begin{cases}
      0, & \textrm{two eyes fail at the same time}\\
      1, & \textrm{the treated eye fails first} \\ 
      2, & \textrm{the untreated eye fails first}. \nonumber
   \end{cases}
\end{eqnarray}
Moreover, the dependent competing risks model based on the family $BVF(\alpha_0, \alpha_1, \alpha_2, \lambda)$ proposed here is an ideal canidate for this data, as there are ties in the data. Out of total 71 observations, there were 28 cases where the treated eye failed first, 33 cases where the untreated eye failed first, and 10 cases where failure occurred in both eyes at the same time.   

The members of the family $BVF(\alpha_0, \alpha_1, \alpha_2, \lambda)$ were fitted to this data following the approach discussed in Section \ref{sec:model-inference}; the results of model fitting for the three models are given in Table \ref{DRS}.      

\begin{figure}
\centering
\subfloat{\includegraphics[width=.5\linewidth, height=.5\linewidth]{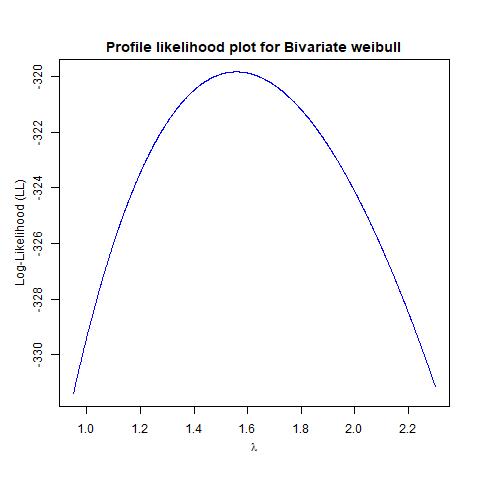}}
\subfloat{\includegraphics[width=.5\linewidth, height=.5\linewidth]{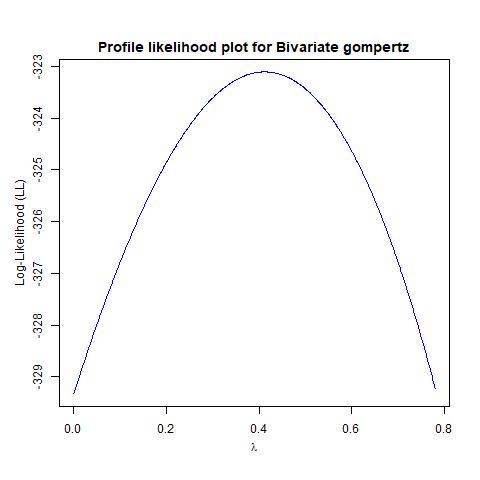}}
\hfill
\subfloat{\includegraphics[width=.5\linewidth, height=.5\linewidth]{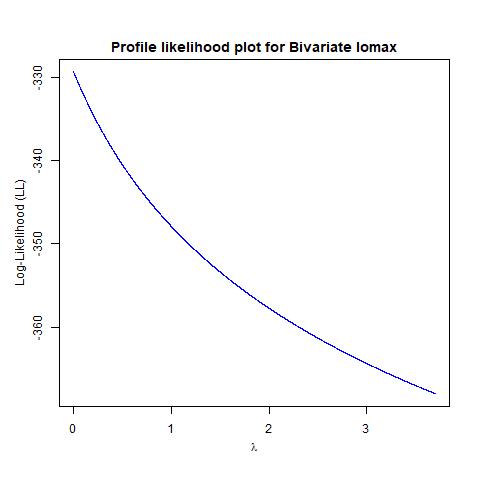}}
\caption{Plot of the profile log-likelihood $p(\lambda)$ for the candidate models based on the Diabetic Retinopathy Data}
\label{diabetic-profile-ll}
\end{figure}

Note that the bivariate Weibull model is the most appropriate model for this data, as it has the largest maximized likelihood ($-319.82$) among the candidate models. The MLEs of parameters do not exist for the bivariate Lomax model based on this data. Indeed, from the plots of the profile log-likelihood $p(\lambda)$ in the parameter $\lambda$ given in Figure \ref{diabetic-profile-ll}, we observe that the profile log-likelihood in case of bivariate Lomax distribution based on the Diabetic Retinopathy data is a monotonic decreasing function.    
\begin{table}[h] 
	\centering 
	\caption{Estimates of parameters for the different members of the family $BVF(\alpha_0, \alpha_1, \alpha_2, \lambda)$ based on the Diabetic Retinopathy Data}
	\begin{tabular}{{|c|c|c|c|c|c|}}
		\toprule
		{\makecell{Member}} &
		{Maximized Likelihood} &
		{\makecell{Parameter Estimates \\ ($\widehat{\alpha}_0$, $\widehat{\alpha}_1$, $\widehat{\alpha}_2$, $\widehat{\lambda}$)}} \\
		\midrule
		\multirow{1}{*} {\makecell{Bivariate Weibull}}
		 				& -319.82 & 0.066, 0.185, 0.218, 1.558 \\	
		\hline
		\multirow{1}{*} {\makecell{Bivariate Gompertz}}
		 				& -323.10 & 0.140, 0.393, 0.463, 0.412 \\
		\hline
		\multirow{1}{*} {\makecell{Bivariate Lomax}}
						& NA & NA \\
		\hline		
	\end{tabular} \label{DRS}
\end{table}

\begin{figure}
\centering
\includegraphics[width=.5\linewidth, height=.5\linewidth]{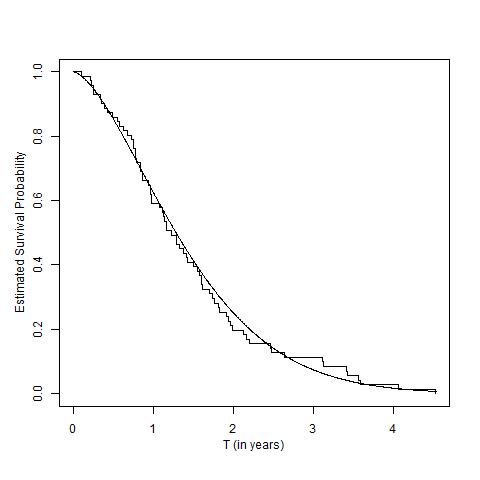}
\caption{Comparison of the fitted bivariate Weibull survival function (the smooth curve) with the Kaplan-Meier curve (the step curve) for the Diabetic Retinopathy Data}
\label{KM-Wei-compare}
\end{figure}

It is of course of interest to see how closely the bivariate Weibull model fits to the data, compared to the nonparametric Kaplan-Meier survival curve, ignoring the cause of failure; the plot is given in Figure \ref{KM-Wei-compare}. It is clear that the two survival curves, parametric and nonparametric, are quite close. 

\section{Conclusion}\label{sec:concl}
In this article, a general approach for modelling dependent competing risks data with a general bivariate family of distributions is presented, including the construction of the bivariate family of distributions, its use in modelling dependent competing risks data, likelihood inference, and related details for the proposed model, and a simple yet powerful model selection approach. Through a detailed Monte Carlo simulation study, it is observed that all the proposed methods of inference in this paper perform quite well. Analysis of a real data is presented as an illustration. 

In summary, this work provides a comprehensive inferential framework for
modelling dependent competing risks data with ties using a general family of bivariate distributions. The model and inferential framework are expected to accommodate dependent competing risks data arising from different spheres of science due to their general and comprehensive nature.

\section*{\sc Acknowledgements} 
The research of Ayon Ganguly is supported by the Mathematical Research Impact
Centric Support (File no.~MTR/2017/000700) from the Science and
Engineering Research Board, Department of Science and Technology, Government of
India. \\
Debanjan Mitra thanks Indian Institute of Management Udaipur for financial
support to carry out this research.

\section*{\sc Declaration of Conflict of Interests} 
The Authors declare that there is no conflict of interest.

\singlespacing

\end{document}